\documentclass[conference]{IEEEtran}
\IEEEoverridecommandlockouts

\usepackage{cite}
\usepackage{url}
\usepackage{amsmath,amssymb,amsfonts}
\usepackage{algorithmic}
\usepackage{graphicx}
\usepackage{textcomp}
\usepackage{xcolor}
\usepackage{bm}
\usepackage{booktabs}
\usepackage{ulem}
\usepackage{tikz}
\usetikzlibrary{arrows.meta,positioning,fit,backgrounds,calc}
\graphicspath{{figs/}{./}}
\def\BibTeX{{\rm B\kern-.05em{\sc i\kern-.025em b}\kern-.08em
T\kern-.1667em\lower.7ex\hbox{E}\kern-.125emX}}

\newcommand{\bX}{\mathbf{X}}
\newcommand{\bw}{\mathbf{w}}
\newcommand{\btheta}{\bm{\theta}}

\newcommand{\Dset}{\mathcal{D}}
\newcommand{\Real}{\mathbb{R}}

\begin{document}

\title{B-ExposNet: Bayesian Deep Learning for EMF Exposure
Mapping with Uncertainty Quantification
}

\author{\IEEEauthorblockN{
Yukun Liu\IEEEauthorrefmark{1},   
Shanshan Wang\IEEEauthorrefmark{1},  
Paul Lagouanelle\IEEEauthorrefmark{1},  
Yarui Zhang\IEEEauthorrefmark{2}, 
Joe Wiart\IEEEauthorrefmark{1}
}                                   
\IEEEauthorblockA{\IEEEauthorrefmark{1}
Chair C2M, LTCI, Télécom Paris, Polytechnic Institute of Paris, Palaiseau, France }
\IEEEauthorblockA{\IEEEauthorrefmark{2}
SATIE, ENS Paris-Saclay, CNRS, Université Paris-Saclay, Gif-sur-Yvette, France \\
Corresponding Author: yukun.liu@telecom-paris.fr
}
}
\maketitle

\begin{abstract}
Electromagnetic field exposure in urban environments is highly variable, and conventional exposure maps provide a deterministic estimate, without considering its uncertainty. To address the gap, this paper proposes B-ExposNet, a Bayesian neural network for spatial exposure estimation with uncertainty quantification. B-ExposNet extends ExposNet by treating the weights of regression head as random variables, inferred by variational inference, so that the model provides probabilistic estimates from posterior predictive distribution. The method is evaluated on drive-test measurement campaigns conducted in three French cities, and compared with conventional method such as universal Kriging, in both predictive accuracy and the quality of the estimated uncertainty. On the test region, B-ExposNet achieves a mean absolute percentage error of 28.19\% and a continuous ranked probability score of 0.198~V/m, improving on canonical methods in both accuracy and uncertainty quantification.

\end{abstract}


\section{Introduction}
The densification of mobile networks and the continued expansion of network infrastructure have raised public concern about electromagnetic field (EMF) exposure. The International Commission on Non-Ionizing Radiation Protection published guidelines for compliance assessment \cite{icnirp2020} in response to public concerns.
Although current research has not established adverse health effects from EMF exposure below guideline limits, EMF exposure remains an environmental factor that requires continuous monitoring. However, measurement campaigns are costly and provide only limited spatial coverage, making measurement-only monitoring impractical for an entire urban area \cite{wang2025comprehensive}. Reliable models are therefore needed to estimate exposure where direct measurements are unavailable.

This need has motivated the development of data-driven methods for EMF exposure estimation. Deep learning, in particular, has been increasingly adopted, offering a data-driven approach to learning the non-linear relationships between environmental inputs and exposure levels. In \cite{emf_ann}, an artificial neural network performs spatial extrapolation with features manually extracted from public datasets, while \cite{exposnet} proposes a convolutional architecture that takes city images as environmental input and thereby avoids explicit feature engineering. Recurrent models have also been applied to temporal exposure forecasting \cite{emf_temp}. These studies establish the applicability of deep learning to EMF exposure prediction.

One aspect that remains insufficiently addressed is the variability of exposure, resulting from user activity and from the complexity of the surrounding urban environment. A single deterministic prediction provides only a partial description of EMF exposure, making uncertainty quantification as important as estimating the exposure level itself.

As deep learning has become prevalent, uncertainty quantification (UQ) techniques adapted to neural architectures have emerged alongside it. Monte Carlo dropout randomly deactivates neurons at each forward pass, generating multiple stochastic outputs whose dispersion is used to estimate predictive uncertainty \cite{dropout}. Deep ensembles train several networks from independent initializations and quantify uncertainty from the variability among their predictions \cite{de}. Bayesian neural networks (BNN) treat the weights as random variables whose posterior distribution is inferred from the observed data \cite{bnn_intro}. 

Nevertheless, these UQ techniques remains insufficiently explored for spatial EMF exposure estimation. This paper therefore proposes an uncertainty-aware framework for spatial EMF exposure estimation, named B-ExposNet. Building on the BNN approach, B-ExposNet keeps the convolutional backbone of ExposNet \cite{exposnet} deterministic and replaces its regression head by Bayesian layers. This hybrid architecture preserves the predictive capability of the original model and additionally provides location-dependent uncertainty estimates. B-ExposNet is trained and evaluated on drive-test measurements collected in Lyon, Massy and Paris, and compared with universal kriging \cite{kriging} in terms of point-prediction accuracy, interval sharpness, empirical coverage and probabilistic predictive performance.



The remainder of this paper is organized as follows. Section~\ref{sec:method} formulates the spatial EMF exposure estimation problem, presents the B-ExposNet architecture and defines the evaluation metrics. Section~\ref{sec:exp} describes the measurement campaigns and the data preparation. Section~\ref{sec:results} reports the experimental results and compares B-ExposNet with a conventional benchmark method. Section~\ref{sec:concl} summerizes the main findings and concludes the paper.

\section{Methodology}\label{sec:method}

\subsection{Problem Formulation}\label{sec:problem}
Let $\Omega \subset \Real^{2}$ denote the spatial domain, $\ell \in \Omega$ denote an arbitrary location, $y(\ell) \in \Real_{+}$ denote the exposure level at $\ell$, in V/m, and let $\bX(\ell)$ collect the auxiliary environmental and network information associated with that location. The objective is to learn a predictive framework $\mathcal{M}$ that maps the auxiliary information at any location to an exposure estimate $\hat{y}(\ell)$ together with an associated uncertainty estimate $\hat{u}(\ell) \in \Real_{+}$:
\begin{equation}
\bigl( \hat{y}(\ell),\, \hat{u}(\ell) \bigr) = \mathcal{M}\bigl( \bX(\ell) \bigr)
\label{eq:framework}
\end{equation}

This formulation defines the general spatial exposure-estimation with uncertainty quantification problem. 
Under a data-driven approach, $\mathcal{M}$ is learned from a training set of $N$ locations, $\Dset = \{ ( \bX(\ell_i),\, y_i ) \}_{i=1}^{N}$, with the measured levels $y_i = y(\ell_i)$ as targets. 

\subsection{Environmental Feature Engineering}\label{sec:features}

The design of the context tensor $\bX(\ell)$ follows \cite{exposnet}. Each location $\ell$ is described by a multi-channel image patch:
\begin{equation}
\bX(\ell) \in \Real^{C \times H \times W}
\label{eq:input}
\end{equation}
obtained by rasterizing a square context window centered on $\ell$: the center pixel corresponds to the prediction location, while the surrounding pixels encode the local environmental and physical context. Here $C$ is the number of input channels and $H \times W$ the pixel resolution where each channel is rasterised, and the resolution determines how finely the local context is resolved. In practice, $C = 11$ and $H = W = 128$. The eleven channels \cite{exposnet} comprise 
\begin{itemize}
\item two trajectory-related maps --- the drive-test trajectory and the trajectory--antenna relation map; \item two environmental maps --- land cover and building height; \item seven frequency-specific antenna-influence maps, for the 700, 800, 900, 1800, 2100, 2600 and 3500~MHz bands, derived from the locations, azimuths and types of the declared base-station antennas.
\end{itemize}

The geographic and network layers are built from the French National Institute of Geographic and Forest Information (IGN) \cite{ign_data} and the ANFR Cartoradio database \cite{anfr_cartoradio}.

\subsection{Bayesian Formulation}\label{sec:bnn}

The surrogate model $\mathcal{M}$ of \eqref{eq:framework} is realized as a Bayesian neural network. Let $f_{\bw}$ denote the network mapping a context tensor to a scalar E-field level, where $\bw$ collects the weights treated as random variables; the remaining weights are deterministic and are suppressed from the notation. In contrast to a deterministic regressor, the BNN defines a probability distribution over the exposure, normally through a likelihood that places a Gaussian \sout{of} variance $\sigma_{n}^{2}$ around its output:
\begin{equation}
p\bigl( y \mid \bX(\ell), \bw \bigr) =
\mathcal{N}\bigl( y ;\, f_{\bw}( \bX(\ell) ),\, \sigma_{n}^{2} \bigr)
\label{eq:likelihood}
\end{equation}
so that $p(\Dset \mid \bw) = \prod_{i} p( y_i \mid \bX(\ell_i), \bw )$ measures how well the weights $\bw$ explain the observed exposure levels.

Rather than fitting a single value of $\bw$, the Bayesian formulation assigns a prior $\pi(\bw)$ to the weights and updates it with the evidence provided by the data through Bayes' rule:
\begin{equation}
p\bigl( \bw \mid \Dset \bigr) =
\frac{p\bigl( \Dset \mid \bw \bigr)\, \pi(\bw)}{p(\Dset)}
\label{eq:posterior}
\end{equation}
The prediction at a new location $\ell$ is then not a single output but the posterior predictive distribution obtained by marginalizing the weights:
\begin{equation}
p\bigl( y \mid \bX(\ell), \Dset \bigr) = \int
p\bigl( y \mid \bX(\ell), \bw \bigr)\, p\bigl( \bw \mid \Dset \bigr)\,
\mathrm{d}\bw
\label{eq:postpred}
\end{equation}

The estimation problem \eqref{eq:framework} is thereby transformed into the characterization of this distribution, whose expectation provides the exposure estimate and whose standard deviation provides the associated uncertainty:
\begin{equation}
\begin{aligned}
\hat{y}(\ell) &= \mathbb{E}\bigl[\, y \mid \bX(\ell), \Dset \,\bigr]  \\[2pt]
\hat{u}(\ell) &= \sqrt{ \mathrm{Var}\bigl[\, y \mid \bX(\ell), \Dset \,\bigr] } 
\end{aligned}
\label{eq:moments}
\end{equation}

\subsection{B-ExposNet}
The rest of this section describes the architecture of B-ExposNet, the variational approximation, and the Monte Carlo evaluation.

\begin{figure*}
\centering
\input{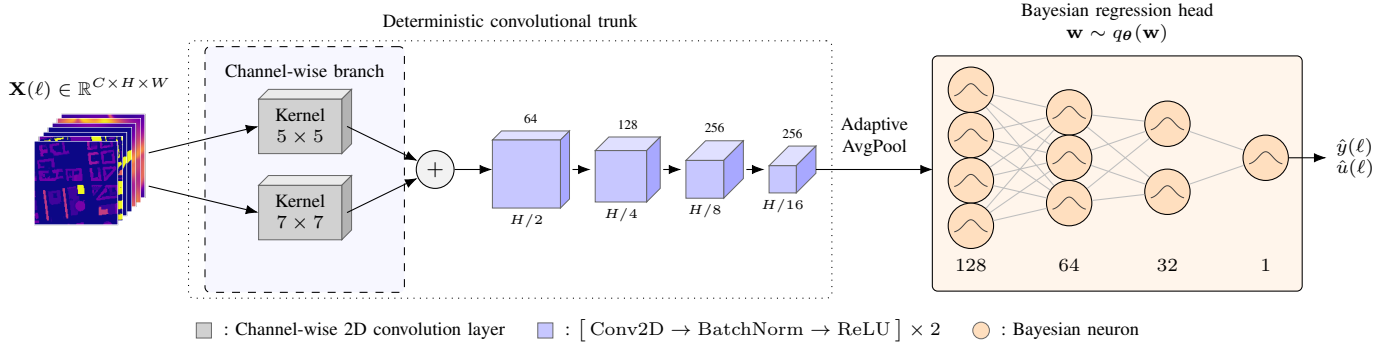}
\caption{B-ExposNet architecture. {The convolutional trunk on the left is deterministic; Bayesian regression head on the right contains weights  $\bw$ as random variables, with posterior approximated by $q_{\btheta}(\bw)$.}}
\label{fig:arch}
\end{figure*}

\subsubsection{Architecture}

The predictive model follows the ExposNet architecture \cite{exposnet}, and is illustrated in Fig.~\ref{fig:arch}. Each of the $C = 11$ input channels is first processed by a 2D convolution layer with parallel $5 \times 5$ and $7 \times 7$ kernels, so that fine local variations and broader spatial patterns are captured separately for each heterogeneous input map. The channel-wise representations are then concatenated and encoded by a sequence of convolution--BatchNormalization--ReLU blocks interleaved with max pooling, which expand the channel dimension while reducing the spatial resolution, so that deeper layers integrate context from a progressively larger neighborhood of $\ell$. Adaptive average pooling finally produces a feature vector that a fully connected head $128 \rightarrow 64 \rightarrow 32 \rightarrow 1$ maps to the E-field level at the central location. In B-ExposNet this head becomes the Bayesian head shown on the right of Fig.~\ref{fig:arch}, whose neurons are probabilistic.

Concretely, $\bw$ collects the weights of the $128 \rightarrow 64 \rightarrow 32$ part of the bayesian head; all remaining weights of the network are deterministic. Variational inference over the convolutional kernels would be computationally demanding and unstable to train, whereas this partially stochastic design keeps the training cost close to that of the deterministic model: the convolutional trunk acts as a learned embedding of the environmental context, on top of which a Bayesian regressor operates.

\subsubsection{Variational inference}
The posterior \eqref{eq:posterior} is approximated by a mean-field Gaussian $q_{\btheta}(\bw) = \prod_{k} \mathcal{N}( w_k ; \mu_k, \sigma_k^{2} )$, with variational parameters $\btheta = \{ (\mu_k, \sigma_k) \}_k$. The variational parameters are obtained by minimizing the negative evidence lower bound (ELBO):
\begin{equation}
\mathcal{L}(\btheta) =
\mathrm{KL}\bigl[\, q_{\btheta}(\bw) \,\|\, \pi(\bw) \,\bigr]
- \mathbb{E}_{q_{\btheta}(\bw)}
\bigl[\, \log p\bigl( \Dset \mid \bw \bigr) \,\bigr]
\label{eq:elbo}
\end{equation}
where $\mathrm{KL}$ denotes the Kullback--Leibler divergence \cite{kl}. The two terms balance proximity to the prior against fidelity to the data. The objective and its gradients are estimated using Monte Carlo weight samples and the reparameterization procedure employed in Bayes-by-Backprop \cite{bnn}.

\subsubsection{Predictive distribution}
After the variational posterior \(q_{\btheta}(\bw)\) has been learned, it is used to construct the posterior predictive distribution at an unobserved location. Specifically, to estimate the exposure at a new location \(\ell\), \(T\) weight samples \(\bw^{(t)}\sim q_{\btheta}(\bw)\) are propagated through the network, yielding $\hat{y}^{(t)}(\ell) = f_{\bw^{(t)}}\bigl( \bX(\ell) \bigr)$, whose mean and standard deviation are:
\begin{align}
\hat{y}(\ell)
&= \frac{1}{T} \sum_{t=1}^{T} \hat{y}^{(t)}(\ell) 
\nonumber \\[2pt]
\hat{u}(\ell)
&= \sqrt{\frac{1}{T-1} \sum_{t=1}^{T}
\Bigl( \hat{y}^{(t)}(\ell) - \hat{y}(\ell) \Bigr)^{2}}
\label{eq:predstd}
\end{align}
This is the Monte Carlo approximation of \eqref{eq:postpred}, with $q_{\btheta}$ in place of the exact posterior.
Consequently, the estimates in \eqref{eq:predstd} approximate the predictive moments in \eqref{eq:moments} and provide a practical probabilistic realization of the surrogate \(\mathcal{M}\).


\subsection{Evaluation Metrics}\label{sec:metrics}

To evaluate the model performance, point-prediction accuracy, interval quality, and probabilistic accuracy are assessed jointly over the \(N_{\mathrm{t}}\) test locations using the metrics defined below.

\paragraph{Point-prediction accuracy}
The root mean squared error (RMSE) and the mean absolute percentage error (MAPE) are:
\begin{align}
\mathrm{RMSE} &= \sqrt{ \frac{1}{N_{\mathrm{t}}} \sum_{i=1}^{N_{\mathrm{t}}} \bigl( y_i - \hat{y}_i \bigr)^{2} } \nonumber \\[2pt]
\mathrm{MAPE} &= \frac{100}{N_{\mathrm{t}}} \sum_{i=1}^{N_{\mathrm{t}}} \left| \frac{y_i - \hat{y}_i}{y_i} \right| \label{eq:mape}
\end{align}

\paragraph{Interval quality}
Under the Gaussian assumption, the predictive distribution is summarized by an interval $I_i(\tau)$ at nominal coverage level $\tau$:
\begin{equation}
    I_i(\tau) = \bigl[\, \hat{y}_i - z_{(1+\tau)/2}\,\hat{u}_i ,\;\;
    \hat{y}_i + z_{(1+\tau)/2}\,\hat{u}_i \,\bigr]
    \label{eq:interval}
\end{equation}
where $z_{(1+\tau)/2}$ is the standard-normal quantile at level $(1+\tau)/2$, equal to $1.96$ for $\tau = 0.95$. Validity and sharpness are measured by the empirical coverage $\widehat{\mathrm{Cr}}(\tau)$ and the mean prediction interval width (MPIW) at nominal level $\tau$:
\begin{align}
\widehat{\mathrm{Cr}}(\tau) &= \frac{1}{N_{\mathrm{t}}} \sum_{i=1}^{N_{\mathrm{t}}} \mathbb{I}\bigl( y_i \in I_i(\tau) \bigr) \label{eq:cov} \\[2pt]
\mathrm{MPIW}(\tau) &= \frac{1}{N_{\mathrm{t}}} \sum_{i=1}^{N_{\mathrm{t}}} 2 z_{(1+\tau)/2}\, \hat{u}_i \label{eq:mpiw}
\end{align}
The empirical coverage measures the proportion of observations that fall inside the predicted intervals and should be close to the nominal level $\tau$ for well-calibrated model, whereas the MPIW measures the sharpness of those intervals. The two are complementary: a predictive distribution is preferred when it attains the smallest MPIW at a coverage close to the nominal level.

\paragraph{Probabilistic accuracy}
The continuous ranked probability score (CRPS) \cite{matheson1976scoring} evaluates the whole predictive distribution against the observation:
\begin{equation}
\mathrm{CRPS}(F_i, y_i) = \int_{-\infty}^{+\infty}
\bigl[ F_i(t) - \mathbb{I}(t \geq y_i) \bigr]^{2}\, \mathrm{d}t
\label{eq:crps}
\end{equation}
where $F_i$ is the predictive cumulative distribution function; for a Gaussian predictive distribution a closed form is available: 
\begin{equation}
\mathrm{CRPS}(F_i,y_i) = \hat{u}_i \Bigl[ \xi_i \bigl( 2\Phi(\xi_i) - 1 \bigr) + 2\varphi(\xi_i) - \tfrac{1}{\sqrt{\pi}} \Bigr]
\end{equation}
with $\xi_i = (y_i - \hat{y}_i)/\hat{u}_i$, where $\Phi$ and $\varphi$ are the standard-normal cumulative distribution and density. Lower values indicate predictive distributions that are both well centred and appropriately dispersed. 

\section{Experimental Setup}\label{sec:exp}


\subsection{Measurement Campaigns}

The data used in this work were collected during drive-test campaigns conducted in three French cities: Lyon, Massy and Paris. The vehicle moves at an average speed of approximately 30~km/h. The hardware setup comprised a Tektronix RSA306B real-time spectrum analyzer interfaced with a tri-axial antenna probe, an RF switch, and an on-board computer for data logging. The probe was positioned on top of the vehicle for isotropic E-field measurement over a frequency range of 9~kHz to 6.2~GHz. 
To capture the full three-dimensional vector, the switch sequentially toggled between the three orthogonal axes of the antenna, routing the measurement from each axis to the real-time spectrum analyzer for processing. 
A more comprehensive description of the calibration and measurement protocols can be found in \cite{exposnet}.


The routes were planned to encompass a diverse range of urban and suburban environments across the three cities. The total trajectory lengths span approximately on 64~km in Lyon, 26~km in Massy and 22~km in Paris, and traverse residential, commercial and industrial districts as well as public facilities and open spaces. The three campaigns contain 19\,062 E-field measurement samples in Lyon, 6\,373 in Massy and 7\,516 in Paris.

\begin{figure}
\centerline{\includegraphics[width=1.1\linewidth]{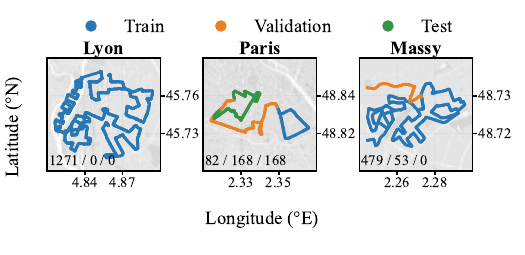}}
\caption{Training / validation / test partition of the sample locations, with the per-city counts given below each sub-figure.}
\label{fig:splits}
\end{figure}	

\subsection{Data Preparation and Splits}
The preprocessing follows \cite{exposnet}. Drive-test measurements are dense along the routes and adjacent samples are strongly correlated, so individual samples are not used directly as targets. Instead, spatial-support centers are placed every 50~m along each route, giving 1\,271 centers in Lyon, 532 in Massy and 418 in Paris, and the target at each center is the root-mean-square of all drive-test measurements falling inside a square window of side 400~m centered on it.

The centers are assigned to the training, validation and test sets in their original order along the routes, without random point-level shuffling, so that held-out segments remain geographically contiguous and neighboring centers are not split across subsets. As shown in Fig.~\ref{fig:splits}, the partition is deliberately asymmetric across cities: all 1\,271 Lyon centers are used for training, Massy contributes 479 training and 53 validation centers, and Paris is divided into 82 training, 168 validation and 168 test centers. The resulting dataset therefore contains 1\,832 training, 221 validation and 168 test instances, and the evaluation is carried out on a contiguous portion of the Paris campaign.

\section{Results and Discussion}\label{sec:results}

B-ExposNet was implemented in PyTorch and trained on an NVIDIA A100 GPU for 50 epochs using Adam optimizer with a learning rate of $10^{-4}$, a weight decay of $10^{-3}$ and a batch size of 64, the ELBO being estimated from three stochastic weight samples per batch. The checkpoint achieving the best validation score was retained for the evaluation. The prior over the stochastic weights is zero-mean Gaussian, $\pi(\bw) = \mathcal{N}(\mathbf{0}, \sigma_{p}^{2}\mathbf{I})$, with $\sigma_{p} = 0.1$. At inference, $T = 100$ stochastic forward passes per location give the predictive mean and standard deviation of \eqref{eq:predstd}. All intervals are reported at the nominal level $\tau = 0.95$, for which $z_{1-\alpha/2} = 1.96$ and \eqref{eq:interval} becomes $\hat{y}(\ell) \pm 1.96\,\hat{u}(\ell)$.

\begin{table}
\caption{Model Performance.}
\label{tab:bnn}
\centering
\begin{tabular}{lcc}
\toprule
\textbf{Metric} & \textbf{B-ExposNet} & \textbf{Kriging} \\
\midrule
RMSE (V/m)          & {0.357} & 0.670 \\
MAPE (\%)           & {28.19}  & 80.01  \\
\midrule
Coverage (\%)       & 79.17  & 92.26 \\
MPIW (V/m)          & 0.854  & 2.194 \\
CRPS (V/m)          & {0.1979} & 0.3902 \\
\bottomrule
\end{tabular}
\end{table}

Table~\ref{tab:bnn} reports B-ExposNet on the Paris test region, against universal Kriging with an exponential covariance. B-ExposNet reaches an RMSE of 0.357~V/m and a MAPE of 28.19\%, against 0.670~V/m and 80.01\% for universal Kriging. Kriging relies on measurements and their spatial correlation alone for prediction, whereas B-ExposNet exploits the environmental and network maps. The gap between the two therefore indicates that spatial correlation alone is not sufficient to reconstruct the exposure field at this resolution, and that the urban structure and the network configuration are necessary inputs.

Kriging requires sparse measurements as inputs, which are expensive to acquire. While BNN uses pure environmental inputs. 

\begin{figure}
\centerline{\includegraphics[width=\columnwidth]{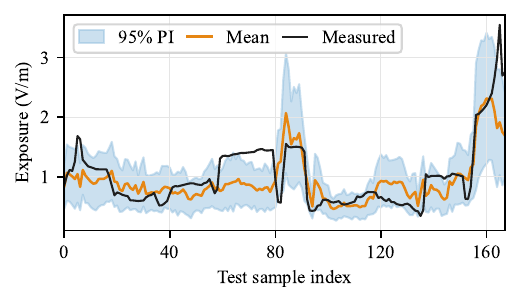}}
\caption{Predictive mean and nominal 95\% interval against the measured exposure on the test route.}
\label{fig:envelope}
\end{figure}

%

In terms of uncertainty quantification, Kriging reaches an empirical coverage of 92.26\%, above the 79.17\% of B-ExposNet, but does so with intervals $2.6$ times wider, its MPIW reaching 2.194~V/m against 0.854~V/m. Coverage and MPIW are complementary as either can be improved at the expense of the other by widening or narrowing the intervals. The CRPS scores the full predictive distribution against the observation and therefore penalizes miscalibration and lack of sharpness jointly, and B-ExposNet attains a CRPS of 0.1979~V/m against 0.3902~V/m for Kriging, which indicates that it provides the better-resolved predictive distribution of the two.

The predictive envelope in Fig.~\ref{fig:envelope} tracks the principal spatial pattern of the measurements and the width of the envelope remains stable over most of the route. It widens markedly around samples~80 and~160, where the model is least confident, and these are also the locations at which the measured exposure is highest. The posterior spread therefore increases in the high-exposure regions, which are the ones least represented in the training set. 

\begin{figure}
\centerline{\includegraphics[width=\columnwidth]{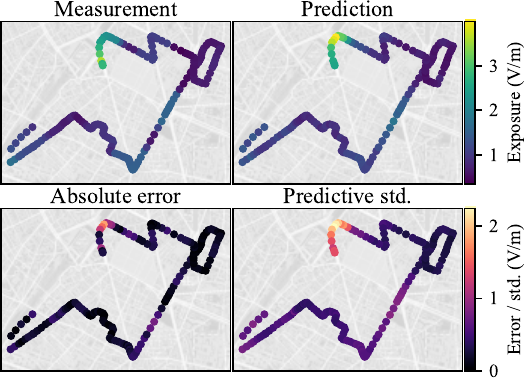}}

\caption{B-ExposNet outputs over the test region. Top: measured exposure (left) and predicted exposure (right). Bottom: absolute error (left) and predictive standard deviation $\hat{\sigma}$ (right).}
\label{fig:spatial}
\end{figure}

Fig.~\ref{fig:spatial} provides a complementary spatial view of the model performance. It maps the measured exposure, predicted exposure, absolute error, and predictive standard deviation along the test route. The predicted exposure reproduces the principal spatial variations observed in the measurements: moderate exposure levels are generally well recovered, whereas the highest values tend to be underestimated. Consequently, the largest errors are concentrated at a limited number of high-exposure locations rather than being uniformly distributed along the route, indicating that the overall RMSE is strongly influenced by these samples. The predictive standard deviation exhibits a similar spatial pattern, remaining low over well-predicted segments and increasing in regions associated with larger errors. This spatial correspondence suggests that B-ExposNet can identify locations at which its predictions are less reliable, providing useful information for prioritizing complementary measurements.

\section{Conclusion}\label{sec:concl}
This paper introduced B-ExposNet, a Bayesian extension of ExposNet for uncertainty-aware spatial EMF exposure estimation. The convolutional feature extractor remains deterministic, as in the original ExposNet architecture, while a mean-field Gaussian posterior is learned over the weights of the regression head using Bayes-by-Backprop. EMF exposure is subsequently estimated from the posterior predictive distribution, which provides both a point prediction and a location-dependent measure of uncertainty.

The proposed framework was validated using real-world drive-test measurements. On the test set, B-ExposNet achieved an RMSE of 0.357~V/m, a MAPE of 28.19\%, and a CRPS of 0.1979~V/m, outperforming universal Kriging in both point-prediction accuracy and probabilistic scoring. The predictive uncertainty was spatially associated with regions of larger error, providing useful information for prioritizing additional measurements. However, the empirical coverage of 79.17\% for the nominal 95\% interval reveals substantial under-coverage. Future work will investigate validation-based interval recalibration and input-dependent aleatoric uncertainty.

\bibliographystyle{IEEEtran}
\bibliography{ref}

@ARTICLE{exposnet,
  author={Zhang, Yarui and Wang, Shanshan and Wiart, Joe},
  journal={IEEE Transactions on Machine Learning in Communications and Networking}, 
  title={{ExposNet}: A Deep Learning Framework for {EMF} Exposure Prediction in Complex Urban Environments}, 
  year={2026},
  volume={4},
  number={},
  pages={1229-1243},
  doi={10.1109/TMLCN.2026.3714431}
}

@article{icnirp2020,
  author    = {{International Commission on Non-Ionizing Radiation Protection}},
  title     = {Guidelines for Limiting Exposure to Electromagnetic Fields (100 {kHz} to 300 {GHz})},
  journal   = {Health Physics},
  volume    = {118},
  number    = {5},
  pages     = {483--524},
  year      = {2020},
  doi       = {10.1097/HP.0000000000001210},
}

@ARTICLE{emf_ann,
  author={Chikha, Wassim Ben and Wang, Shanshan and Wiart, Joe},
  journal={IEEE Access}, 
  title={An Extrapolation Approach for {RF-EMF} Exposure Prediction in an Urban Area Using Artificial Neural Network}, 
  year={2023},
  volume={11},
  number={},
  pages={52686-52694},
  doi={10.1109/ACCESS.2023.3280125}}

@inproceedings{bnn,
    author = {Blundell, Charles and Cornebise, Julien and Kavukcuoglu, Koray and Wierstra, Daan},
    title = {Weight uncertainty in neural networks},
    year = {2015},
    publisher = {JMLR.org},
    booktitle = {Proceedings of the 32nd International Conference on Machine Learning - Volume 37},
    pages = {1613–1622},
    numpages = {10},
    location = {Lille, France},
series = {ICML'15}
}

@inproceedings{de,
 author = {Lakshminarayanan, Balaji and Pritzel, Alexander and Blundell, Charles},
 booktitle = {Advances in Neural Information Processing Systems},
 editor = {I. Guyon and U. Von Luxburg and S. Bengio and H. Wallach and R. Fergus and S. Vishwanathan and R. Garnett},
 pages = {},
 publisher = {Curran Associates, Inc.},
 title = {Simple and Scalable Predictive Uncertainty Estimation using Deep Ensembles},
 volume = {30},
 year = {2017}
}

@article{dropout,
    author = {Srivastava, Nitish and Hinton, Geoffrey and Krizhevsky, Alex and Sutskever, Ilya and Salakhutdinov, Ruslan},
    title = {Dropout: a simple way to prevent neural networks from overfitting},
    year = {2014},
    issue_date = {January 2014},
    publisher = {JMLR.org},
    volume = {15},
    number = {1},
    issn = {1532-4435},
    journal = {J. Mach. Learn. Res.},
    month = jan,
    pages = {1929–1958},
    numpages = {30}
}

@ARTICLE{emf_temp,
  author={Pala, Zeydin},
  journal={IEEE Canadian Journal of Electrical and Computer Engineering}, 
  title={Examining {EMF} Time Series Using Prediction Algorithms With {R}}, 
  year={2021},
  volume={44},
  number={2},
  pages={223-227},
  doi={10.1109/ICJECE.2020.3037805}
}

@article{kriging,
  title={Kriging or polynomial interpolation procedures},
  author={Matheron, G},
  journal={CIMM Transactions},
  volume={70},
  number={1},
  pages={240--244},
  year={1967}
}

@Inbook{bnn_intro,
author="Goan, Ethan
and Fookes, Clinton",
editor="Mengersen, Kerrie L.
and Pudlo, Pierre
and Robert, Christian P.",
title="Bayesian Neural Networks: An Introduction and Survey",
bookTitle="Case Studies in Applied Bayesian Data Science: CIRM Jean-Morlet Chair, Fall 2018",
year="2020",
publisher="Springer International Publishing",
address="Cham",
pages="45--87",
isbn="978-3-030-42553-1",
doi="10.1007/978-3-030-42553-1_3"
}

@misc{ign_data,
  author = {{Institut national de l'information géographique et forestière (IGN)}},
  title = {{Portail IGN: Données géographiques et forestières}},
  year = {2026},
  url = {https://www.ign.fr/},
}

@misc{anfr_cartoradio,
  author = {{Agence nationale des fréquences (ANFR)}},
  title = {{Cartoradio: Carte des antennes et des mesures d'exposition}},
  year = {2026},
  url = {https://www.cartoradio.fr/},
}

@article{matheson1976scoring,
  author  = {Matheson, James E. and Winkler, Robert L.},
  title   = {Scoring Rules for Continuous Probability Distributions},
  journal = {Management Science},
  volume  = {22},
  number  = {10},
  pages   = {1087--1096},
  year    = {1976},
  doi     = {10.1287/mnsc.22.10.1087}
}

@article{wang2025comprehensive,
  title={Comprehensive Measurement-Based Assessment of Downlink RF-EMF Exposure in Urban Environments: Multi-Method Analysis and Intercomparison},
  author={Wang, Shanshan and Zhang, Yarui and Liu, Yukun and Liu, Jiang and Conil, Emmanuelle and Jawad, Ourouk and Samaras, Theodoros and Ourak, Lamine and Wiart, Joe},
  journal={Bioelectromagnetics},
  volume={46},
  number={8},
  pages={e70033},
  year={2025},
  publisher={Wiley Online Library}
}

@article{kl,
  author  = {Kullback, Solomon and Leibler, Richard A.},
  title   = {On information and sufficiency},
  journal = {The Annals of Mathematical Statistics},
  volume  = {22},
  number  = {1},
  pages   = {79--86},
  year    = {1951},
  publisher = {Institute of Mathematical Statistics},
  doi     = {10.1214/aoms/1177729694}
}

\end{document}